\documentclass[aip, apl, reprint]{revtex4-2}
\usepackage{graphicx}
\usepackage{times}
\usepackage{amsmath}
\usepackage{threeparttable}
\usepackage[hidelinks]{hyperref}
\usepackage{float}

\newcommand{\WmK}{W~m$^{-1}$~K$^{-1}$}
\newcommand{\MWmK}{MW~m$^{-2}$~K$^{-1}$}
\newcommand{\MJmK}{MJ~m$^{-3}$~K$^{-1}$}
\newcommand{\micron}{$\mu$m}
\newcommand{\degree}{$^{\circ}$}

\begin{document}

\title{Lock-in infrared thermography: phase analysis for rapid, wide-range thermal conductivity measurements}

\author{Ethan A. Scott}
\affiliation{Department of Mechanical and Aerospace Engineering, University of Virginia, Charlottesville, Virginia 22904, USA}

\author{Jeffrey L. Braun}
\affiliation{Laser Thermal Analysis, Inc., 937 2nd St SE, Charlottesville, VA 22902, USA}

\author{Jessica Reyes}
\affiliation{Department of Mechanical and Aerospace Engineering, University of Virginia, Charlottesville, Virginia 22904, USA}

\author{Bruce Bolliger}
\affiliation{Element Six Technologies, Santa Clara, CA 95054, USA}

\author{Terrence Soares}
\affiliation{Element Six Technologies, Santa Clara, CA 95054, USA}

\author{John T. Gaskins}
\affiliation{Laser Thermal Analysis, Inc., 937 2nd St SE, Charlottesville, VA 22902, USA}

\author{Marko J. Tadjer}
\affiliation{US Naval Research Laboratory, Washington, DC 20375, USA}

\author{Patrick E. Hopkins}
\affiliation{Department of Mechanical and Aerospace Engineering, University of Virginia, Charlottesville, Virginia 22904, USA}
\affiliation{Laser Thermal Analysis, Inc., 937 2nd St SE, Charlottesville, VA 22902, USA}
\affiliation{Department of Materials Science and Engineering, University of Virginia, Charlottesville, Virginia 22904, USA}
\affiliation{Department of Physics, University of Virginia, Charlottesville, Virginia 22904, USA}

\begin{abstract}
We report on a phase-based lock-in thermography approach, combined with a multilayered thermal model (often employed in thermoreflectance analysis), to measure the thermal conductivity of bulk materials and layered structures. The spatial distribution of the material's thermal phase is monitored with an infrared camera, which is locked into the frequency of a modulated laser used to heat the material. This phase distribution is then fit with a thermal model, in which properties such as thermal conductivity are extracted as fit parameters. This approach enables non-contact, front-side measurements, which are insensitive to surface roughness. The technique does not strictly require the application of a transducer layer, but we highlight the practical benefits of applying a removable adhesive layer to serve as a near-surface absorber. We demonstrate the efficacy of the method by measuring materials with thermal conductivities that span over three orders of magnitude ($\sim$ 1 \WmK{} to $>$ 2000~\WmK{}). 
\end{abstract}

\maketitle

\section{Introduction}

Material thermal performance is integral to engineering design. From raw construction material to synthetic materials for advanced  technological applications, heat transfer plays a key role. Therefore, knowledge of properties such as thermal conductivity is essential for accurate prediction of the performance of an engineering design. 

There are many approaches to measuring thermal conductivity: contact, non-contact, optothermal, electrothermal, steady-state, transient, etc.\cite{Pfeifer2025Jul} While each approach has distinct advantages and limitations, Braun et al. \cite{Braun2025Jul} recently outlined universal characteristics of an ideal technique. Among the most important of these are (1) simplicity in use and analysis (2) relevance to a wide range of materials and thermal conductivities (3) high-throughput capability (4) accuracy and repeatability (5) localized nature (6) tolerance to sample geometry and surface condition. To this end, a new technique, thermo-optical plane source method (TOPS), was developed. TOPS employs a robust infrared detection scheme which analyzes the magnitude of a material's temperature rise as a function of incident laser power in order to directly measure thermal conductivity. While the approach satisfies many of the aforementioned characteristics, it is best suited for low thermal conductivity materials\cite{Braun2025Jul} (generally materials with $\kappa <$ 100 \WmK{}).

\begin{figure}
  \centering
  \includegraphics[scale = 0.73]{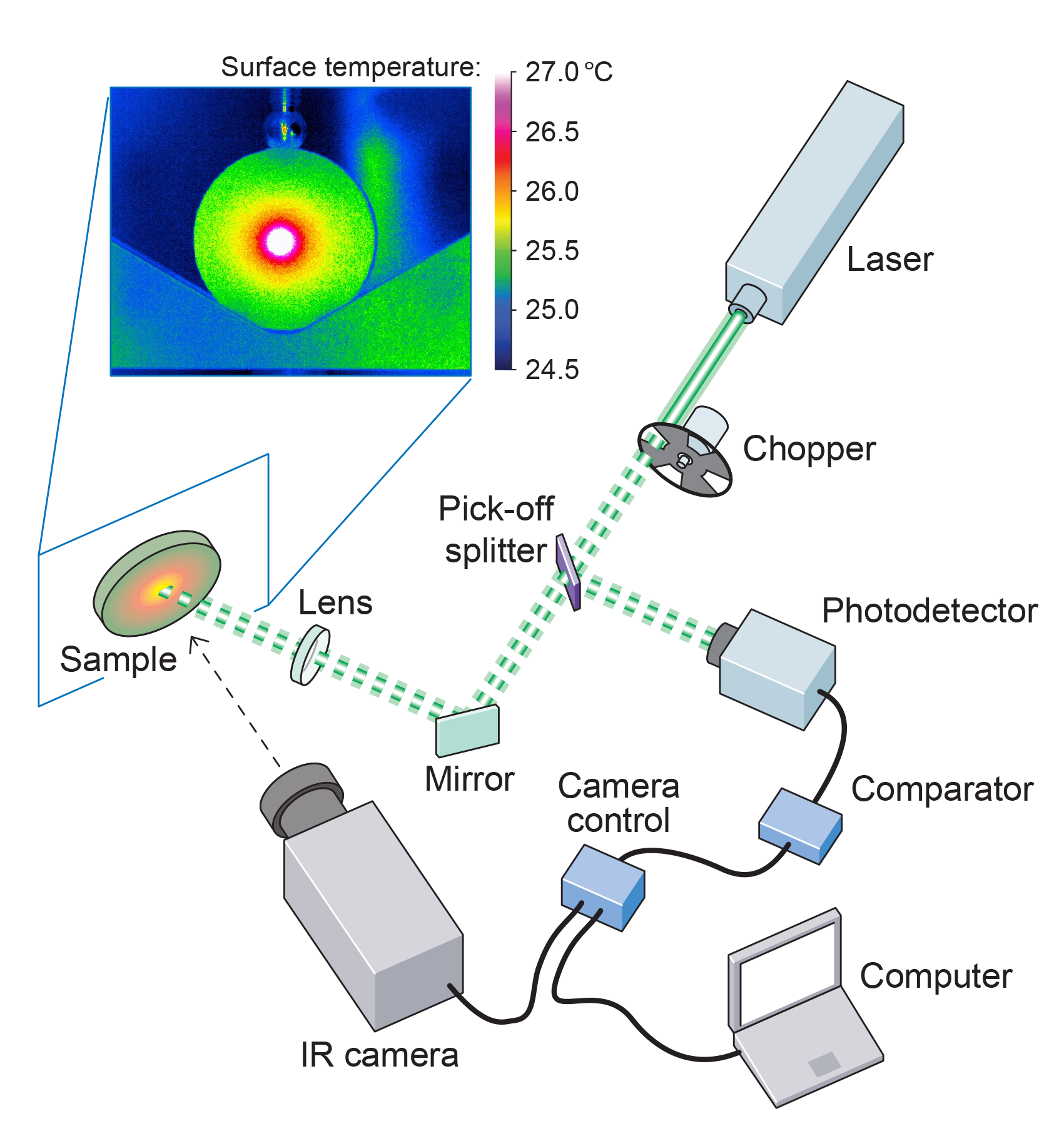}
  \caption{Schematic of experimental setup.}
  \label{fig_system_layout}
\end{figure}

In this work, we combine concepts from the TOPS technique, lock-in infrared thermography (LIT), and continuous-wave thermoreflectance approaches\cite{Schmidt2009Sep, Braun2019Feb}, to enhance the capabilities of non-contact infrared based measurements. This method pushes the upper limit of measurable thermal conductivities while maintaining many of the aforementioned measurement characteristics. The resultant technique is non-contact/non-destructive, requires only single-sided heating/probing access, does not require a polished surface, does not strictly require an absorbing layer/transducer, and can measure high thermal conductivity materials ($\kappa > 2000$ \WmK{}). 

 \begin{figure*}
  \centering
  \includegraphics[scale = 0.4]{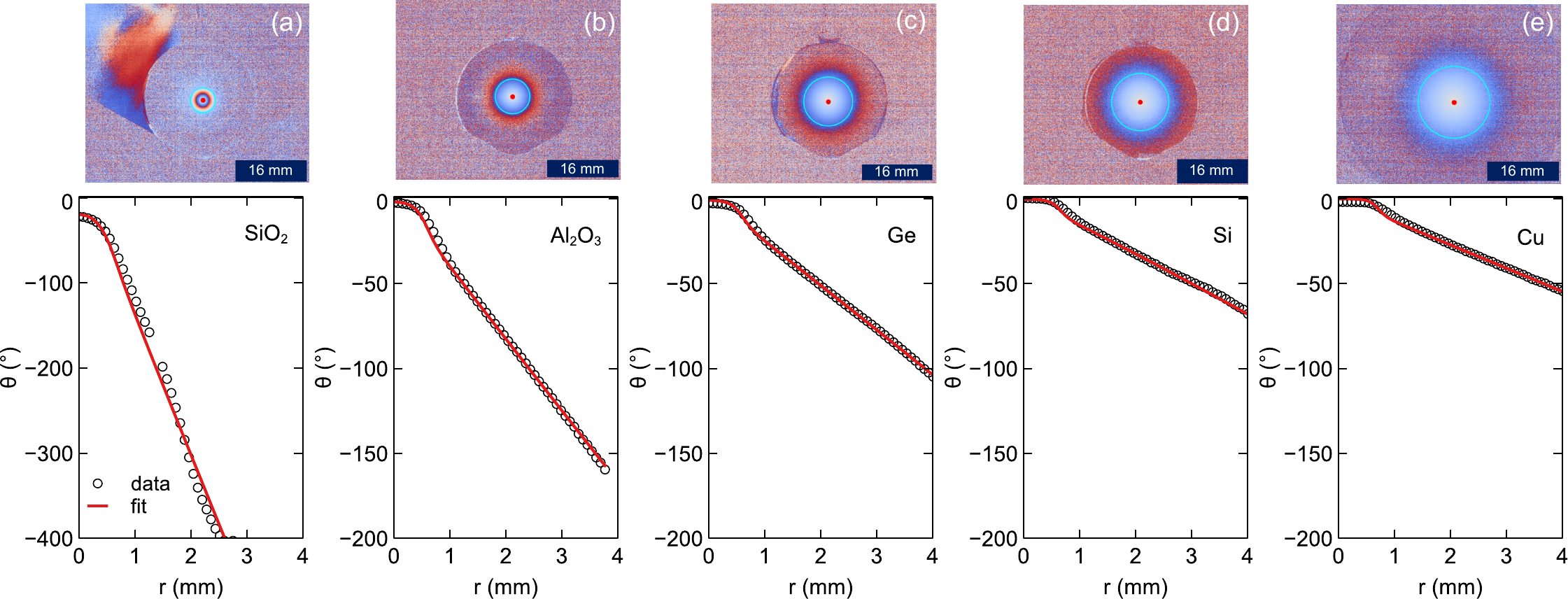}
  \caption{Example thermal phase images, as monitored by the IR camera, for 2 Hz laser modulation. The corresponding radial average of the phase is displayed below each image along with the best fit model for each curve. The blue circles in (a) - (e) denote the cutoff of the fitted radius, $r$.}
  \label{fig_phase_fits}
\end{figure*}

\section{Background and Theory}

There have been numerous implementations of LIT. A few recent examples are demonstrated in Refs. \citenum{Gaitonde2023Jul, Bedoya2023Feb, Bedoya2024Apr, Salazar2025Aug, Cifuentes2017Nov, Kaneko2025Apr, Ishizaki2019Jun}. Despite the variations among these implementations, the common underlying principal of operation is that a material is periodically heated (usually with a modulated laser source) and the thermal response is captured with an infrared (IR) camera and processed via lock-in amplifier to extract the magnitude and phase at the modulation frequency. The data is then analyzed to extract thermal properties. The strength of the technique comes from the lock-in amplification, which enhances the signal-to-noise ratio, allowing for the detection of small temperature variations. A significant benefit of phase analysis is that, in contrast to magnitude, the absorbed power does not need to be known. This reduces the number of model parameters, and in part, is why phase is preferentially modeled in other lock-in based thermal metrologies such as time-domain\cite{Jiang2018Oct} or frequency-domain thermoreflectance\cite{Schmidt2009Sep, Kirsch2024Oct}. 

In this approach, we extract thermal properties from the phase by employing a similar model used for thermoreflectance\cite{Cahill2004Dec, Schmidt2009Sep, Braun2019Feb}. That is, the radial heat diffusion equation is solved for structures in response to a periodic, Gaussian heat source. A benefit of this modeling approach is the ability to accommodate structures with an arbitrary number of layers, thus allowing for consideration of thermo-optic transducer (absorbing) layers or systems which consist of film(s) on a substrate. 

We first solve for the radial- and frequency-dependent temperature profile $T(r,\omega)$ in a similar manner as Refs. \citenum{Braun2019Feb, Braun2017May}. From this, we calculate the thermal phase as $\theta(r,\omega) = \tan^{-1}(Y(r,\omega)/X(r,\omega))$, where $X(r,\omega)$ and $Y(r,\omega)$ are the real and imaginary components of $T(r,\omega)$, respectively. The difference between the modeled and the measured thermal phase is then minimized with a least squares routine in which properties of interest are treated as adjustable parameters.

\section{Experimental Details}

A schematic of the experimental layout is shown in Fig. \ref{fig_system_layout}. The setup consists of a sample, which is periodically heated by a laser and is monitored with a lock-in IR camera. The camera captures the spatial distribution of the thermal phase, which is then output to a computer for refinement and analysis. 

 We utilize a Coherent Verdi V-5 continuous wave laser, which has a central wavelength of 532 nm and maximum optical power output of 5 W. The laser is modulated at a desired frequency with a Stanford Research Systems Model SR542 Precision Optical Chopper. Lenses are used to adjust the laser spot size at the sample surface. We then measure the focused spot size with a Thorlabs BP209-VIS beam profiler. We typically utilize a laser beam with focused $1/e^{2}$ diameter of $\leq$~1~mm and incident powers of $\leq$ 0.5 W, which translates to a peak surface temperature rise on the order of several degrees.

The thermal phase of the sample is monitored with an InfraTec ImageIR 8380 IR camera, which is a mid-wave IR camera that has a 640 $\times$ 512 pixel resolution and $\leq$ 20 mK temperature resolution. We lock into the heating frequency of the sample by diverting a small percentage of the laser with a beam sampler (Thorlabs BSF10-A), and direct it into a photodetector (Thorlabs DET10A). The DET10A output is converted into a transistor-transistor logic (TTL) compatible signal with a comparator (Pulse Research Lab PRL-350TTL). This is then supplied as the reference signal for the IR camera with a camera control module. The control module in Fig. \ref{fig_system_layout} represents a combination of the InfraTec Active Thermography Control Module and BreakOut-Box. The thermal phase is acquired over a specified time or number of periods, and output into a tabulated data file, which is then analyzed with a thermal model. We generally collect 100 periods per measurement using a camera frame rate of 200 Hz.

\section{Results and Discussion}

\subsection{Sample measurements}

\begin{figure*}
  \centering
  \includegraphics[scale = 0.5]{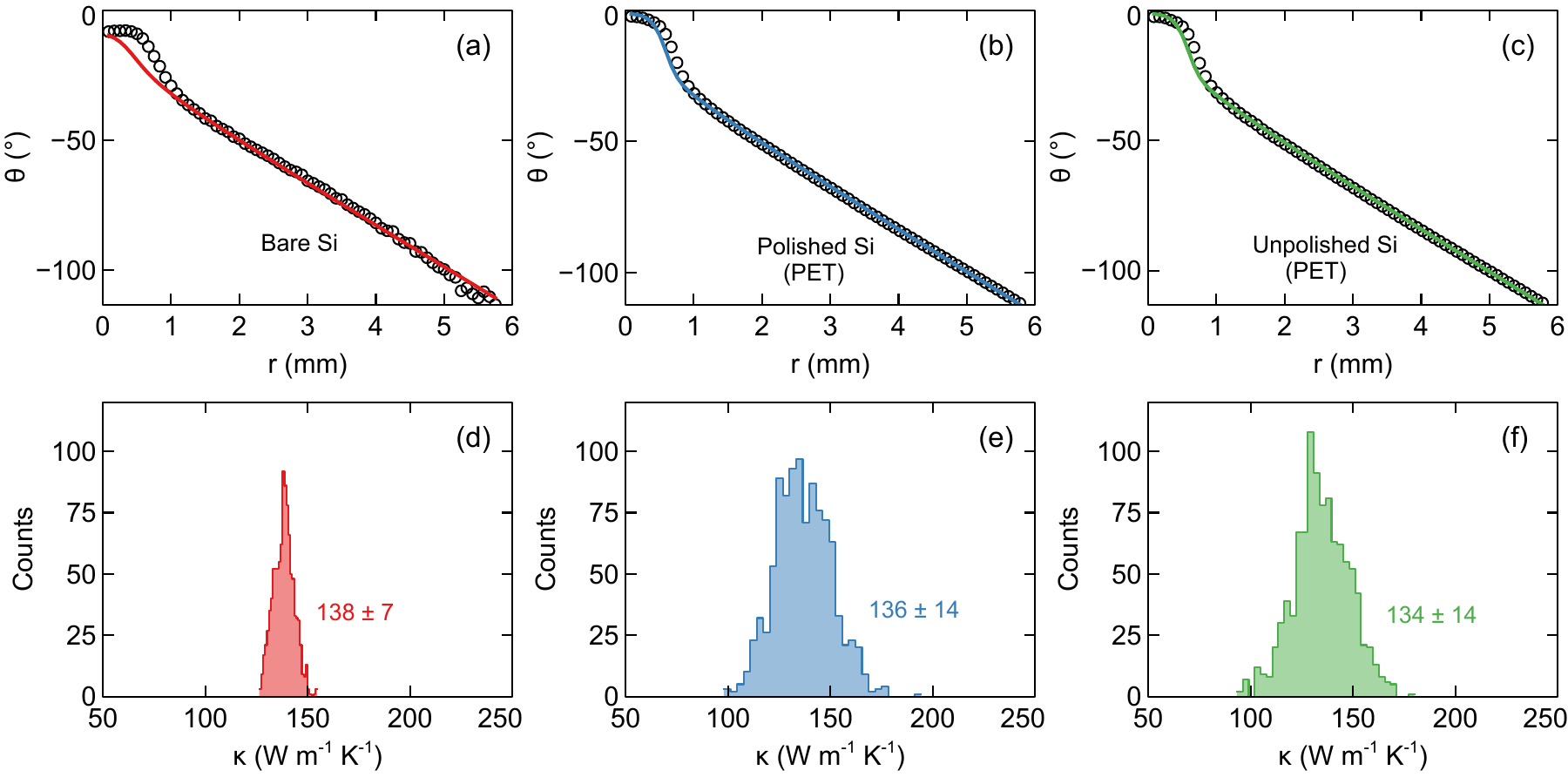}
  \caption{Fitted phase data from a bare Si wafer (a), Si wafer with PET on the polished side (b), and a Si wafer with PET on the unpolished side (c). (d) - (f) show associated histograms of the fitted thermal conductivity from Monte Carlo simulation.}
  \label{fig_Si_surface_examples}
\end{figure*}

The spatially distributed thermal phase is radially symmetric for a material heated with a circular (Gaussian) beam. To minimize variation in the data that occurs from sample imperfections, we take a radial average of the phase for a vector that extends outward from the center of the laser heated region to the edge of a user-defined region of interest (ROI), and is rotated 360 degrees about the center point. The center point corresponds to the point of least negative phase. We use a peak-finding algorithm to automate center point determination. The radius of the ROI is typically on the order of millimeters, but is a function of the laser spot size and sample properties. For example, the spatial extent of the lateral temperature rise is higher for more conductive materials. Fig. \ref{fig_phase_fits}(a-e) demonstrates this, showing the spatially distributed phase and the defined ROI for a variety of samples. The radially averaged phase is shown directly beneath the corresponding contour image. 

All of the phase plots in Fig. \ref{fig_phase_fits} show a symmetric profile about $r$ = 0 mm, in which the phase monotonically decreases as $r$ is increased. This indicates that the temperature rise becomes increasingly out of phase relative to the frequency of the heating laser as position from the origin of the heating source is increased. For a given heating frequency, this offset in phase is more pronounced for materials with lower thermal conductivity. This is shown by the slopes of the curves in \ref{fig_phase_fits} where each panel shows a material with successively increasing thermal conductivity. If the slope is sufficiently large, additional data processing is required. For example, the camera records the thermal phase for -180\degree{} $\leq \theta \leq$  180 \degree{}. If the thermal phase is lower than this threshold, the phase is wrapped, and must be manually corrected by subtracting 180 \degree{} from the apparent value; data points within the discontinuity of the phase wrap are discarded. An example of this is shown for the SiO$_2$ sample of Fig. \ref{fig_phase_fits}. 

We note that the plots in Fig. \ref{fig_phase_fits} show the thermal phase for materials coated with a thin, optically absorbent adhesive layer. But, to be clear, this layer is not strictly necessary if the irradiated material is absorptive of the incident laser power. However, absorbing layers have been demonstrated in prior reports\cite{Braun2025Jul} as a means to improve absorption of the incident beam, which ultimately reduces the laser output power requirement. This is particularly useful for materials that are highly thermally conductive, optically transmissive, or reflective of the heating laser. In those cases, we use a black polymer-based absorber, which is a 5 micron thick tape consisting of a black polyethylene terephthalate (PET) film with an acrylic adhesive. While this adds an extra step, this transducer application is faster than the metalization required for thermoreflectance and is more easily removable by peeling or dissolving it in acetone.

\begin{figure*}
  \centering
  \includegraphics[scale = 0.5]{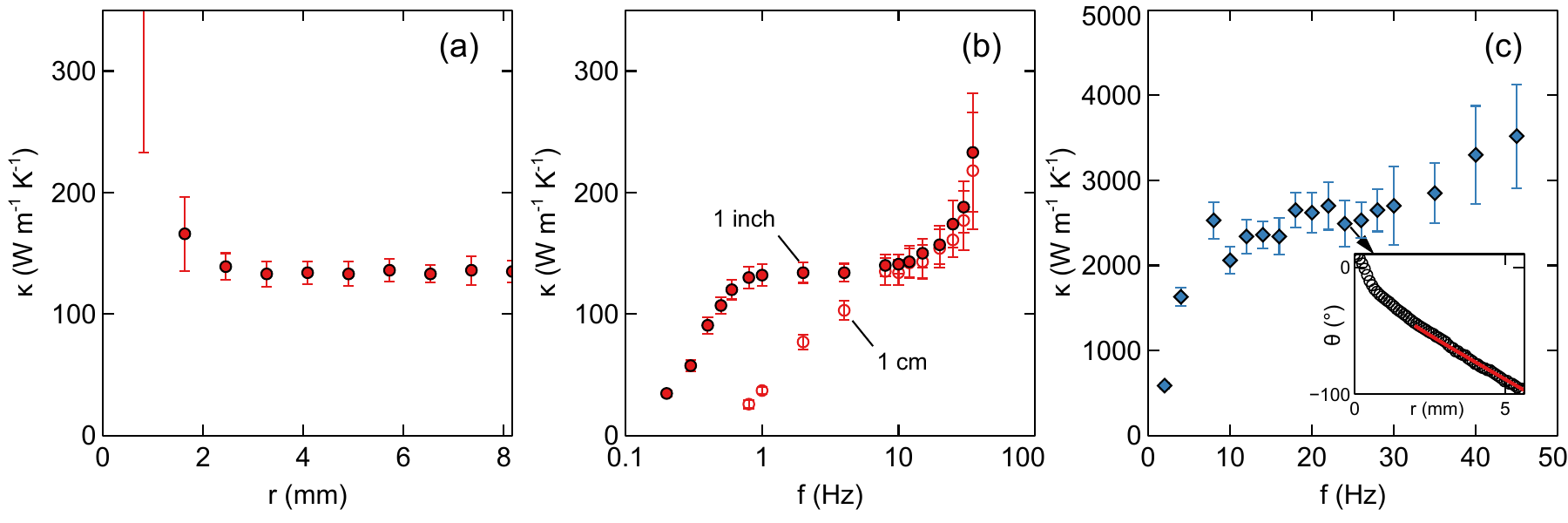}
  \caption{Measurement parameter effects on fitted thermal conductivity. (a) Shows the effect of fit cutoff radius. (b) and (c) show the effect of modulation frequency for Si (cleaved, 1 inch and 1 cm on a side) and diamond (33 mm, E6 TM220), respectively. The inset of (c) shows an example fit of diamond at 24 Hz.}
  \label{fig_LIT_heuristics}
\end{figure*}

We demonstrate the effect of the absorber layer by comparing the apparent thermal conductivity of a material measured with and without an absorber layer. Specifically, we measure silicon wafers (University Wafer, (100), SSP, P/B 1-100 $\Omega{}$-cm, 724 microns thick) which were cleaved, one inch on a side. LIT measurements were performed on each sample using a laser spot size of 880 microns (1/e$^2$ diameter), chopped at a frequency of 2 Hz. Fig. \ref{fig_Si_surface_examples}(a) and (b) show the radially-averaged phase of a bare silicon wafer and a PET-coated wafer, respectively. For reference, we also measure a piece of the silicon with PET on the unpolished side. Fig. \ref{fig_Si_surface_examples}(d) - (f) shows the fitted thermal conductivity of each sample and the associated uncertainty. In each case, the thermal conductivity is within uncertainty of the other samples, indicating minimal difference in the effect of sample preparation or surface condition. This is attributed to the spatial extent of the measurement. In contrast, themoreflectance generally probes a region several microns in diameter and is tolerant to tens of nanometers of RMS roughness \cite{Jiang2018Oct}. LIT data is averaged over an area of several square millimeters and is tolerant to surface roughness on the order of microns, compatible with unpolished wafers. 

Regarding the uncertainty of the model fit, we apply the same approach used in thermoreflectance. That is, we add in quadrature the uncertainty from the spot-to-spot variation, uncertainty from the least squares model fit (employing a 95\% confidence interval), as well as uncertainty in the fixed model parameters (accounted for with Monte Carlo analysis \cite{Yang2016Jan, Kirsch2024Oct}). For the PET layer, we assume an average thermal conductivity of 0.28 \WmK{} and average heat capacity of 1.62 \MJmK{}, which are representative of thermoplastic polymers \cite{Xie2016Feb}. We note, however, that the thermal properties of this transducer layer have an insignificant effect on the thermal model, evidenced also by the similarity in data between samples with and without the transducer, shown in Fig. \ref{fig_Si_surface_examples}. For the interface between the PET layer and the sample substrate, we assume a low interface conductance of $G$ = 1 \MWmK{}, but again note that there is minimal sensitivity to the assumed input (which we address through sensitivity analysis in Section \ref{subsec:sensitivity}). For Monte Carlo simulation of fixed parameters, we account for 10\% variability in the assumed thermal conductivity, heat capacity, and interface conductance of the PET layer, one micron variability in PET thickness, 10\% uncertainty in the measured beam diameter, and also uncertainty in the measured pixel density. We also account for uncertainty in the thermal absorption depth of the incident laser\cite{Yang2016Mar}. We conservatively account for this by treating it as the entire thickness of the PET film, but observe a negligible effect. Ultimately, we find the uncertainty in the measured camera pixel density to yield the highest source of measurement uncertainty.  

Fig. \ref{fig_Si_surface_examples}(d)-(f) shows the Monte Carlo simulation \cite{Yang2016Jan} result from  1000 fit iterations for the phase models of Fig. \ref{fig_Si_surface_examples}(a) - (c), respectively. The total uncertainty is approximately 10\% for the fitted thermal conductivity, and is slightly lower for bare Si, as there are fewer model parameters. However, despite lower uncertainty in the Monte Carlo analysis, there is higher variability in the measured phase of the bare wafer due to lower absorption, and thus a lower temperature rise. For this reason, a higher incident power was required on the bare Si (2.6 W compared to 0.35 W). 

\subsection{Bounds of Operation}

Outside of thermal model considerations, precautions must be taken with regard to sample preparation and camera configuration. In an ideal scenario, samples would be infinitely large such that there is no spatial limitation to thermal diffusion. Additionally, the camera would have an infinitely high sampling rate to avoid any aliasing effects. In reality, samples are of a finite size, which limits the extent of thermal spreading as well as the range over which the thermal phase can be averaged for analysis. As a result, confinement of heat in small samples can lead to offsets in the thermal phase observed by the IR camera. One solution is to restrict the extent of the thermal spreading within a given period of time. This can be achieved by increasing the frequency of the laser modulation \cite{Wolf2004Dec}. Slow modulation results in greater thermal diffusion length and vice-versa. However, the frame rate of the camera is limited, which imposes practical limits on the modulation frequency. In our case, the frame rate is limited to 200 Hz, which is the product of the LIT frequency (periods per second) and the sampling rate (frames per period). We find that noticeable offsets in the phase occur at sampling rates below $\sim$7 frames per period, and thus we are generally limited for LIT frequencies less than 30 Hz, which also depends upon the particular sample. Use of a higher modulation frequency is preferable as it reduces the measurement duration. However, frequencies above the order of 1 Hz are generally not required unless the sample is highly thermally conductive (hundreds of \WmK{}), or the sample has a low or moderate thermal conductivity and is small (side lengths $\leq$ 1 cm). 

We demonstrate these effects by measuring the apparent thermal conductivity of a sample as a function of fit radius, modulation frequency, and sample size. We perform these measurements on the same silicon used for the tests in Fig. \ref{fig_Si_surface_examples}. Fig \ref{fig_LIT_heuristics}(a) shows the effect of fitted radius. In this case, there is minimal difference in fitted thermal conductivity for $r \geq$ 2 mm. For smaller radii, the thermal model is largely insensitive to the thermal conductivity, which we discuss in further detail in Section \ref{subsec:sensitivity}. 

\begin{table*}[]
\begin{threeparttable}
    
\caption{Measured thermal conductivities compared with literature values. For reference, we also list the heat capacity, thickness, and radius of each sample. The fused silica, sapphire, germanium and silicon samples were purchased from Thorlabs. The copper was purchased from McMaster-Carr, and the diamond was supplied by Element Six.}
\label{table_measured_vs_lit}
\begin{tabular}{lcccccc}
\hline\hline
Sample & Manufacturer No. & $\kappa_{\mathrm{Meas.}}$ (\WmK{}) & $\kappa_{\mathrm{Ref.}}$ (\WmK{}) & $C$ (\MJmK{}) & $d$ (mm) & $r_\mathrm{sample}$ (mm)\\
\hline
Fused Silica & PF10-03 & 1.30 $\pm$ 0.08  & 1.38 \cite{Braun2025Jul}      & 1.63 \cite{Scott2018Nov}     & 6.12    &12.66\\
Sapphire     & WG31050 & 35.5 $\pm$ 1.7   & 33.4\cite{Braun2025Jul}       & 3.06 \cite{Ziade2020Dec}     & 4.85    &12.65\\
Germanium    & WG91050 & 55.0 $\pm$ 2.4   & 55.7 \cite{Braun2025Jul}      & 1.72 \cite{Paterson2020Jun}  & 5.00    &12.69\\
Silicon      & WG81050 & 141 $\pm$ 8     & 148 \cite{Liu2005Jun}          & 1.65 \cite{Scott2018Nov}     & 5.08    &12.67\\
Copper       & 9103K2     & 385 $\pm$ 15     & 398 \cite{Cancellieri2020Nov} & 3.45 \cite{Cancellieri2020Nov} & 12.70 &25.4\\
Diamond      & TM220   & 2490 $\pm$ 310   & 2200\tnote{a}                 & 1.78 \cite{Scott2021Feb}     & 1.27    &16.5\\
\hline\hline
\end{tabular}

\begin{tablenotes}
\item[a]Manufacturer specified value
\end{tablenotes}
\end{threeparttable}
\end{table*}

Fig. \ref{fig_LIT_heuristics}(b) shows the effect of modulation frequency. The laser beam was chopped at frequencies between 0.2 Hz and 50 Hz (which are the limits of the lock-in camera in our configuration). On the low-frequency end (f $<$ 1 Hz), we observe a reduction in the apparent thermal conductivity. We attribute this to confinement of thermal diffusion, which in turn yields an offset in the thermal phase. The effect is exacerbated as the lateral sample dimensions are reduced. For example, Fig. \ref{fig_LIT_heuristics}(b) also shows a silicon sample cleaved into a coupon 1 cm on a side, which yields a further-reduced thermal conductivity for a given modulation frequency below 8 Hz. As the modulation frequency is increased, there is a threshold above which the apparent thermal conductivity remains consistent. If the modulation frequency is further increased, there is an increase in the apparent thermal conductivity, which can be attributed to an artificial offset in the phase due to insufficient frame rate. Above this point, we note that that the model fit becomes increasingly poor and the slope of the phase is not well captured by the model, and gives rise to larger error bars. This effect occurs for f $\geq$ 20 Hz for the Si sample in Fig. \ref{fig_LIT_heuristics}(b).

If samples are sufficiently large (e.g, the 1 inch sample in Fig. \ref{fig_LIT_heuristics}(b)), the region of consistent thermal conductivity is clear, and the fitted values within this region are representative of the true material thermal conductivity. When samples are small and conductive (e.g., the 1 cm Si sample), this region may be narrow or unclear, ultimately necessitating fabrication of a larger sample. As another example, Fig. \ref{fig_LIT_heuristics}(c) shows the same frequency-dependent analysis of a CVD diamond wafer (E6 TM220, 33 mm diameter). The sample is of similar size as the cleaved silicon. However, it is more than an order of magnitude more thermally conductive. Therefore, we observe a modified region of consistent thermal conductivity in the region from 10 Hz - 30 Hz, which averages to approximately 2490 \WmK{}. Above 30 Hz, there is a similar reduction in fit quality due to modification of the apparent phase. In order to measure smaller samples of diamond, a camera with a higher frame rate would be required.

\begin{figure}
  \centering
  \includegraphics[scale = 0.65]{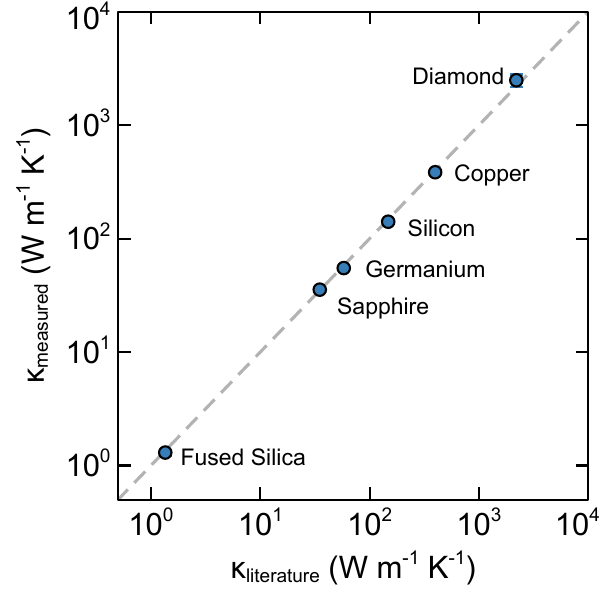}
  \caption{Comparison of measured thermal conductivity values with those expected from the literature for a variety of bulk samples.}
  \label{fig_meas_vs_lit}
\end{figure}

Once the camera and fit parameters are established (in our case, $f\geq$ 1 Hz for most materials and, $r\geq$ 2 mm), accurate measurements can be obtained for materials spanning a wide range of thermal conductivity. In Fig. \ref{fig_meas_vs_lit}, we show measurements for several bulk samples compared against their literature or manufacturer claimed values, tabulated in Table \ref{table_measured_vs_lit}. We note that all measurements in Fig. \ref{fig_meas_vs_lit} (except for the copper and diamond) are optical windows, one inch in diameter, from Thorlabs. The copper sample was purchased from McMaster-Carr (part number 9103K2), and the diamond is a 33 mm diameter wafer of the commercially available TM220. In all cases, a modulation frequency of 2 Hz was used, except for the diamond, which was measured over a range of 10 Hz - 30 Hz, as previously discussed. Ultimately, we find agreement for each material, within uncertainty, of the literature or manufacturer specified values. 

\subsection{Sensitivity considerations}\label{subsec:sensitivity}

\begin{figure*}
  \centering
  \includegraphics[scale = 0.6]{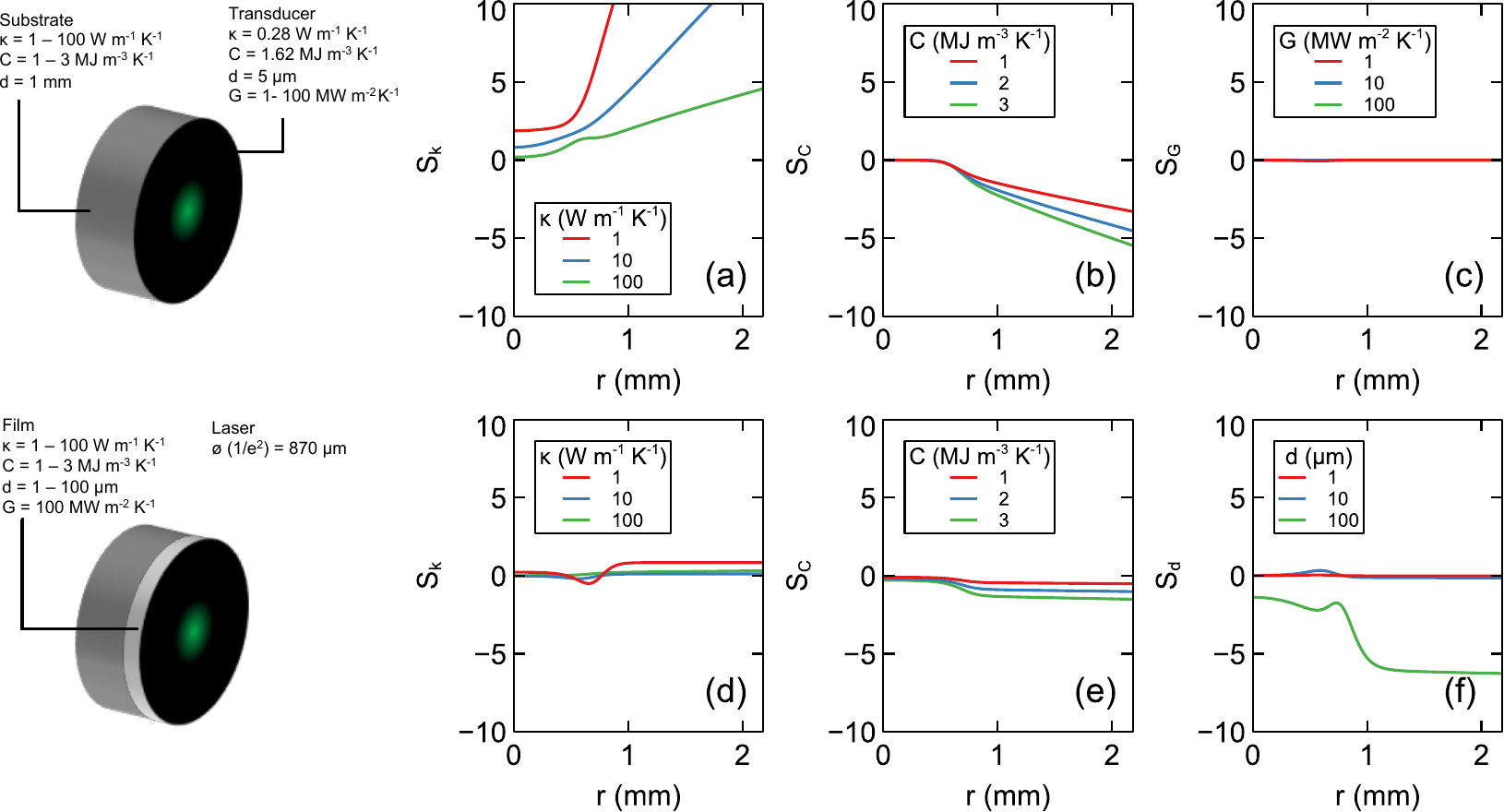}
  \caption{Sensitivity, $S_x$, of the phase to a given parameter, $x$, as a function of radius. Plots (a) - (c) display sensitivities for a transducer/substrate sample stack. In this case, the sample is assumed to have a thermal conductivity, heat capacity, and thermal boundary conductance of 100 \WmK{}, 2 \MJmK, and 10 \MWmK, respectively, unless otherwise noted. Plots (d) - (f) correspond to a transducer/film/substrate stack. The same substrate conditions are assumed, and the film is assumed to have a thermal conductivity, heat capacity, and thickness of 1 \WmK{}, 2 \MJmK, and 50 \micron{}, unless otherwise noted. The front- and backside thermal boundary conductances are assumed to be 10 \MWmK and 100 \MWmK, respectively.}
  \label{fig_sensitivity}
\end{figure*}

Fig. \ref{fig_LIT_heuristics}(a) demonstrates the effect of the fitted region cutoff radius: minimal changes in resultant thermal conductivity for $r\geq{}$2 mm. In the other extreme, there are large errors in the apparent fitted value, which is attributed to negligible sensitivity to the substrate thermal conductivity. That is, the thermal phase is not strongly impacted by a change in apparent thermal conductivity within that region. To quantify sensitivity of the phase, $S$, to a given parameter, $x$, we define $S_x$ in a similar method used by Yang et al. \cite{Yang2013Oct} for FDTR. In this case, we calculate $S_x$ as a function of radius as $S_x(r) = \theta|_{1.1x}(r)-\theta_{0.9x}(r)$, where $\theta|_{1.1x}(r)$ and $\theta|_{0.9x}(r)$ denote the radial dependence of the thermal phase in response to positive and negative 10\% perturbations in the variable of interest, respectively. 

In Fig. \ref{fig_sensitivity} we consider parameter sensitivity for bulk materials (Fig. \ref{fig_sensitivity}(a) - (c)) and for a film on substrate (Fig.\ref{fig_sensitivity}(d) - (f)). In both cases, we assume that a layer of PET has been adhered to the surface. For the bulk material, we consider the sensitivity of the substrate thermal conductivity, heat capacity, and also the interface conductance between the absorber layer and substrate. In each case, we assume a 1 mm thick substrate with default thermal properties of $\kappa$ = 100 \WmK{}, $C$ = 2 \MJmK{}, $G$ = 10 MW m$^{-2}$ K$^{-1}$. The curves in each panel show the effect of perturbing one of these parameters. With regard to substrate thermal conductivity, $S_{\kappa}$ is minimized near $r = 0$, and becomes more pronounced as $r$ increases. Therefore, sensitivity to $\kappa$ scales with $r$, which underscores the benefits of larger samples, particularly for highly thermally conductive samples. This also shows that the entire radius is not needed in order to accurately fit the thermal conductivity. As long as the correct slope is established, it is sufficient to accurately fit the thermal conductivity of a bulk substrate. This approach of fitting the slope of the phase been previously demonstrated \cite{Mendioroz2009Jul, Fabbri1995Jun}, and can be useful if there are defected regions or aberrations in the collected phase at certain radii. For the materials measured in this study, we fit the entire radial range, except for diamond, in which case we fit only the linear region for $r\geq 2$ mm where sensitivity to thermal conductivity is maximized, and any aberrations near $r = 0$ mm can be neglected. An example fit of diamond is shown in the inset of Fig. \ref{fig_LIT_heuristics}(c) for a modulation frequency of 24 Hz.

We note that the thermal conductivity of the materials measured in this study are considered as isotropic when performing a model fit. Thus, the $\kappa_\mathrm{measured}$ value is representative of an isotropic value. However, anisotropic materials could be handled in a similar manner as in thermoreflectance experiments\cite{Ziade2020Dec}. That is, fitting for either in-plane or cross-plane values. We generally find significantly higher sensitivity to the in-plane thermal spreading, and therefore the in-plane thermal conductivity. However, for certain sample geometries such as thin foils or membranes, or for highly anisotropic materials, there can also be sensitivity to cross-plane spreading. 

Sensitivity to heat capacity shows a similar, but negative trend as thermal conductivity. This is because the measurement is fundamentally sensitive to the thermal diffusivity, where thermal conductivity and heat capacity have an opposite scaling effect on the frequency-dependent thermal response\cite{Bedoya2024Apr}. Therefore, the heat capacity and thermal conductivity cannot be simultaneously fit. One parameter or the other must be assumed. We tabulate the heat capacity values assumed in these measurements in Table. \ref{table_measured_vs_lit}. 

We also show the sensitivity to the interface conductance between the PET/substrate in Fig. \ref{fig_sensitivity}(c) for $1 \leq G \leq 100$ MW m$^{-2}$ K$^{-1}$. In short, there is minimal sensitivity to the interface. We assume a comparably low value (e.g. 1 MW m$^{-2}$ K$^{-1}$) for the adhesive/substrate interface relative to the metal transducer/substrate interface conductance found in thermoreflectance samples (which are generally on the order of 10$^1$ - 10$^2$ \MWmK{})\cite{Cheaito2015Jan, Aller2021Jun, Monachon2016Jul}. However, the assumption for $G$ yields minimal impact on the fitted thermal conductivity. The lack of sensitivity can be attributed to the much larger (millimeter) spatial scales of the measurement compared to the microscale heater radius and thermal diffusion depths of thermoreflectance. 

In Fig. \ref{fig_sensitivity}(d)-(f), we show sensitivity calculations for a film on a substrate. For the substrate, we assume the same default thermal properties mentioned previously. For the film, we assume to front- and backside thermal boundary conductances of 10 \MWmK{} and 100 \MWmK{}, respectively. The default thermal conductivity, heat capacity, and thickness of the film is assumed to be 1 \WmK{}, 2 \MJmK{}, and 50 \micron{}, respectively. Compared to the bulk substrate, there is significantly lower sensitivity to the thermal conductivity and heat capacity despite the assumed value. This can again be attributed to the large spatial extent of the measurement's thermal diffusion length relative to the thermal resistance of the film. For this example, the there is no appreciable sensitivity to the film until the thickness is $\geq$ 10s of microns. One approach to enhance the measurement of thin films is by minimizing the substrate thickness (e.g., via polishing or lapping the substrate) until a point where $S_{\kappa_\mathrm{film}} > S_{\kappa_\mathrm{substrate}}$. Alternatively, the substrate may be removed completely, and the film probed directly as a suspended membrane \cite{Jiang2023Sep, Alasli2025Jan}.

\section{Summary}

In summary, we discuss a thermal metrology approach which combines principles of established lock-in IR thermography as well as thermoreflectance analysis, which enables wide-range thermal conductivity measurement while overcoming several experimental limitations. The approach is non-contact, requires only sing-sided sample access (heating and measurements both occur on the front side), and is tolerant of surface roughness. We demonstrate that an absorbing layer is not required, however a thin, removable PET adhesive reduces incident power requirements. We also discuss limitations of the technique, with regard to equipment requirements as well as measurement sensitivity. Ultimately, We validate this approach by measuring an array of materials with thermal conductivities spanning three orders of magnitude, as high as $\kappa >$ 2000 \WmK{}, and find good agreement between measured and literature values.

\section{Acknowledgments}
The authors are grateful to Element Six for providing the diamond wafer for thermal analysis. Research at the Naval Research Laboratory was partially supported by the Office of Naval Research. J.R. and P.E.H. appreciate support from the Office of Naval Research, Grant Number N00014-23-1-2630. E.A.S., J.L.B. and J.T.G. appreciate support from the Department of War through the STTR Program, Grant Number N68335-25-C-0590.

\section{Data Availability}
The data that support the findings of this study are available from the corresponding author upon reasonable request.

\bibliographystyle{apsrev4-2}
\bibliography{refs_LIT}

\begin{thebibliography}{33}%
\makeatletter
\providecommand \@ifxundefined [1]{%
 \@ifx{#1\undefined}
}%
\providecommand \@ifnum [1]{%
 \ifnum #1\expandafter \@firstoftwo
 \else \expandafter \@secondoftwo
 \fi
}%
\providecommand \@ifx [1]{%
 \ifx #1\expandafter \@firstoftwo
 \else \expandafter \@secondoftwo
 \fi
}%
\providecommand \natexlab [1]{#1}%
\providecommand \enquote  [1]{``#1''}%
\providecommand \bibnamefont  [1]{#1}%
\providecommand \bibfnamefont [1]{#1}%
\providecommand \citenamefont [1]{#1}%
\providecommand \href@noop [0]{\@secondoftwo}%
\providecommand \href [0]{\begingroup \@sanitize@url \@href}%
\providecommand \@href[1]{\@@startlink{#1}\@@href}%
\providecommand \@@href[1]{\endgroup#1\@@endlink}%
\providecommand \@sanitize@url [0]{\catcode `\\12\catcode `\$12\catcode
  `\&12\catcode `\#12\catcode `\^12\catcode `\_12\catcode `\%12\relax}%
\providecommand \@@startlink[1]{}%
\providecommand \@@endlink[0]{}%
\providecommand \url  [0]{\begingroup\@sanitize@url \@url }%
\providecommand \@url [1]{\endgroup\@href {#1}{\urlprefix }}%
\providecommand \urlprefix  [0]{URL }%
\providecommand \Eprint [0]{\href }%
\providecommand \doibase [0]{https://doi.org/}%
\providecommand \selectlanguage [0]{\@gobble}%
\providecommand \bibinfo  [0]{\@secondoftwo}%
\providecommand \bibfield  [0]{\@secondoftwo}%
\providecommand \translation [1]{[#1]}%
\providecommand \BibitemOpen [0]{}%
\providecommand \bibitemStop [0]{}%
\providecommand \bibitemNoStop [0]{.\EOS\space}%
\providecommand \EOS [0]{\spacefactor3000\relax}%
\providecommand \BibitemShut  [1]{\csname bibitem#1\endcsname}%
\let\auto@bib@innerbib\@empty
\bibitem [{\citenamefont {Pfeifer}\ \emph {et~al.}(2025)\citenamefont
  {Pfeifer}, \citenamefont {Schonfeld}, \citenamefont {Scott}, \citenamefont
  {Aller}, \citenamefont {Gaskins}, \citenamefont {Olson}, \citenamefont
  {Braun}, \citenamefont {Graham},\ and\ \citenamefont
  {Hopkins}}]{Pfeifer2025Jul}%
  \BibitemOpen
  \bibfield  {author} {\bibinfo {author} {\bibfnamefont {T.~W.}\ \bibnamefont
  {Pfeifer}}, \bibinfo {author} {\bibfnamefont {H.~B.}\ \bibnamefont
  {Schonfeld}}, \bibinfo {author} {\bibfnamefont {E.~A.}\ \bibnamefont
  {Scott}}, \bibinfo {author} {\bibfnamefont {H.~T.}\ \bibnamefont {Aller}},
  \bibinfo {author} {\bibfnamefont {J.~T.}\ \bibnamefont {Gaskins}}, \bibinfo
  {author} {\bibfnamefont {D.~H.}\ \bibnamefont {Olson}}, \bibinfo {author}
  {\bibfnamefont {J.~L.}\ \bibnamefont {Braun}}, \bibinfo {author}
  {\bibfnamefont {S.}~\bibnamefont {Graham}},\ and\ \bibinfo {author}
  {\bibfnamefont {P.~E.}\ \bibnamefont {Hopkins}},\ }\href
  {https://doi.org/10.1146/annurev-matsci-080423-010435} {\bibfield  {journal}
  {\bibinfo  {journal} {Annu. Rev. Mater. Res.}\ ,\ \bibinfo {pages} {37}}
  (\bibinfo {year} {2025})}\BibitemShut {NoStop}%
\bibitem [{\citenamefont {Braun}\ \emph {et~al.}(2025)\citenamefont {Braun},
  \citenamefont {Baines}, \citenamefont {Gaskins},\ and\ \citenamefont
  {Hopkins}}]{Braun2025Jul}%
  \BibitemOpen
  \bibfield  {author} {\bibinfo {author} {\bibfnamefont {J.~L.}\ \bibnamefont
  {Braun}}, \bibinfo {author} {\bibfnamefont {B.~N.}\ \bibnamefont {Baines}},
  \bibinfo {author} {\bibfnamefont {J.~T.}\ \bibnamefont {Gaskins}},\ and\
  \bibinfo {author} {\bibfnamefont {P.~E.}\ \bibnamefont {Hopkins}},\
  }\bibfield  {journal} {\bibinfo  {journal} {Rev. Sci. Instrum.}\ }\textbf
  {\bibinfo {volume} {96}},\ \href {https://doi.org/10.1063/5.0267492}
  {10.1063/5.0267492} (\bibinfo {year} {2025})\BibitemShut {NoStop}%
\bibitem [{\citenamefont {Schmidt}\ \emph {et~al.}(2009)\citenamefont
  {Schmidt}, \citenamefont {Cheaito},\ and\ \citenamefont
  {Chiesa}}]{Schmidt2009Sep}%
  \BibitemOpen
  \bibfield  {author} {\bibinfo {author} {\bibfnamefont {A.~J.}\ \bibnamefont
  {Schmidt}}, \bibinfo {author} {\bibfnamefont {R.}~\bibnamefont {Cheaito}},\
  and\ \bibinfo {author} {\bibfnamefont {M.}~\bibnamefont {Chiesa}},\ }\href
  {https://doi.org/10.1063/1.3212673} {\bibfield  {journal} {\bibinfo
  {journal} {Rev. Sci. Instrum.}\ }\textbf {\bibinfo {volume} {80}},\ \bibinfo
  {pages} {094901} (\bibinfo {year} {2009})}\BibitemShut {NoStop}%
\bibitem [{\citenamefont {Braun}\ \emph {et~al.}(2019)\citenamefont {Braun},
  \citenamefont {Olson}, \citenamefont {Gaskins},\ and\ \citenamefont
  {Hopkins}}]{Braun2019Feb}%
  \BibitemOpen
  \bibfield  {author} {\bibinfo {author} {\bibfnamefont {J.~L.}\ \bibnamefont
  {Braun}}, \bibinfo {author} {\bibfnamefont {D.~H.}\ \bibnamefont {Olson}},
  \bibinfo {author} {\bibfnamefont {J.~T.}\ \bibnamefont {Gaskins}},\ and\
  \bibinfo {author} {\bibfnamefont {P.~E.}\ \bibnamefont {Hopkins}},\
  }\bibfield  {journal} {\bibinfo  {journal} {Rev. Sci. Instrum.}\ }\textbf
  {\bibinfo {volume} {90}},\ \href {https://doi.org/10.1063/1.5056182}
  {10.1063/1.5056182} (\bibinfo {year} {2019})\BibitemShut {NoStop}%
\bibitem [{\citenamefont {Gaitonde}\ \emph {et~al.}(2023)\citenamefont
  {Gaitonde}, \citenamefont {Candadai}, \citenamefont {Weibel},\ and\
  \citenamefont {Marconnet}}]{Gaitonde2023Jul}%
  \BibitemOpen
  \bibfield  {author} {\bibinfo {author} {\bibfnamefont {A.~U.}\ \bibnamefont
  {Gaitonde}}, \bibinfo {author} {\bibfnamefont {A.~A.}\ \bibnamefont
  {Candadai}}, \bibinfo {author} {\bibfnamefont {J.~A.}\ \bibnamefont
  {Weibel}},\ and\ \bibinfo {author} {\bibfnamefont {A.~M.}\ \bibnamefont
  {Marconnet}},\ }\href {https://doi.org/10.1063/5.0149659} {\bibfield
  {journal} {\bibinfo  {journal} {Rev. Sci. Instrum.}\ }\textbf {\bibinfo
  {volume} {94}},\ \bibinfo {pages} {074904} (\bibinfo {year}
  {2023})}\BibitemShut {NoStop}%
\bibitem [{\citenamefont {Bedoya}\ \emph {et~al.}(2023)\citenamefont {Bedoya},
  \citenamefont {Mar{\ifmmode\acute{\imath}\else\'{\i}\fi}n}, \citenamefont
  {Puld{\ifmmode\acute{o}\else\'{o}\fi}n},\ and\ \citenamefont
  {Garc{\ifmmode\acute{\imath}\else\'{\i}\fi}a-Segundo}}]{Bedoya2023Feb}%
  \BibitemOpen
  \bibfield  {author} {\bibinfo {author} {\bibfnamefont {A.}~\bibnamefont
  {Bedoya}}, \bibinfo {author} {\bibfnamefont {E.}~\bibnamefont
  {Mar{\ifmmode\acute{\imath}\else\'{\i}\fi}n}}, \bibinfo {author}
  {\bibfnamefont {J.~J.}\ \bibnamefont
  {Puld{\ifmmode\acute{o}\else\'{o}\fi}n}},\ and\ \bibinfo {author}
  {\bibfnamefont {C.}~\bibnamefont
  {Garc{\ifmmode\acute{\imath}\else\'{\i}\fi}a-Segundo}},\ }\href
  {https://doi.org/10.1007/s10765-022-03138-2} {\bibfield  {journal} {\bibinfo
  {journal} {Int. J. Thermophys.}\ }\textbf {\bibinfo {volume} {44}},\ \bibinfo
  {pages} {27} (\bibinfo {year} {2023})}\BibitemShut {NoStop}%
\bibitem [{\citenamefont {Bedoya}\ \emph {et~al.}(2024)\citenamefont {Bedoya},
  \citenamefont {Salazar}, \citenamefont {Mendioroz},\ and\ \citenamefont
  {Mar{\ifmmode\acute{\imath}\else\'{\i}\fi}n}}]{Bedoya2024Apr}%
  \BibitemOpen
  \bibfield  {author} {\bibinfo {author} {\bibfnamefont {A.}~\bibnamefont
  {Bedoya}}, \bibinfo {author} {\bibfnamefont {A.}~\bibnamefont {Salazar}},
  \bibinfo {author} {\bibfnamefont {A.}~\bibnamefont {Mendioroz}},\ and\
  \bibinfo {author} {\bibfnamefont {E.}~\bibnamefont
  {Mar{\ifmmode\acute{\imath}\else\'{\i}\fi}n}},\ }\bibfield  {journal}
  {\bibinfo  {journal} {J. Appl. Phys.}\ }\textbf {\bibinfo {volume} {135}},\
  \href {https://doi.org/10.1063/5.0198882} {10.1063/5.0198882} (\bibinfo
  {year} {2024})\BibitemShut {NoStop}%
\bibitem [{\citenamefont {Salazar}\ and\ \citenamefont
  {Mendioroz}(2025)}]{Salazar2025Aug}%
  \BibitemOpen
  \bibfield  {author} {\bibinfo {author} {\bibfnamefont {A.}~\bibnamefont
  {Salazar}}\ and\ \bibinfo {author} {\bibfnamefont {A.}~\bibnamefont
  {Mendioroz}},\ }\href {https://doi.org/10.1016/j.ijthermalsci.2025.109928}
  {\bibfield  {journal} {\bibinfo  {journal} {Int. J. Therm. Sci.}\ }\textbf
  {\bibinfo {volume} {214}},\ \bibinfo {pages} {109928} (\bibinfo {year}
  {2025})}\BibitemShut {NoStop}%
\bibitem [{\citenamefont {Cifuentes}\ \emph {et~al.}(2017)\citenamefont
  {Cifuentes}, \citenamefont {Mendioroz},\ and\ \citenamefont
  {Salazar}}]{Cifuentes2017Nov}%
  \BibitemOpen
  \bibfield  {author} {\bibinfo {author} {\bibfnamefont
  {{\ifmmode\acute{A}\else\'{A}\fi}.}~\bibnamefont {Cifuentes}}, \bibinfo
  {author} {\bibfnamefont {A.}~\bibnamefont {Mendioroz}},\ and\ \bibinfo
  {author} {\bibfnamefont {A.}~\bibnamefont {Salazar}},\ }\href
  {https://doi.org/10.1016/j.ijthermalsci.2017.07.023} {\bibfield  {journal}
  {\bibinfo  {journal} {Int. J. Therm. Sci.}\ }\textbf {\bibinfo {volume}
  {121}},\ \bibinfo {pages} {305} (\bibinfo {year} {2017})}\BibitemShut
  {NoStop}%
\bibitem [{\citenamefont {Kaneko}\ \emph {et~al.}(2025)\citenamefont {Kaneko},
  \citenamefont {Fujita}, \citenamefont {Ishizaki},\ and\ \citenamefont
  {Nagano}}]{Kaneko2025Apr}%
  \BibitemOpen
  \bibfield  {author} {\bibinfo {author} {\bibfnamefont {Y.}~\bibnamefont
  {Kaneko}}, \bibinfo {author} {\bibfnamefont {R.}~\bibnamefont {Fujita}},
  \bibinfo {author} {\bibfnamefont {T.}~\bibnamefont {Ishizaki}},\ and\
  \bibinfo {author} {\bibfnamefont {H.}~\bibnamefont {Nagano}},\ }\href
  {https://doi.org/10.1007/s10765-025-03523-7} {\bibfield  {journal} {\bibinfo
  {journal} {Int. J. Thermophys.}\ }\textbf {\bibinfo {volume} {46}},\ \bibinfo
  {pages} {50} (\bibinfo {year} {2025})}\BibitemShut {NoStop}%
\bibitem [{\citenamefont {Ishizaki}\ and\ \citenamefont
  {Nagano}(2019)}]{Ishizaki2019Jun}%
  \BibitemOpen
  \bibfield  {author} {\bibinfo {author} {\bibfnamefont {T.}~\bibnamefont
  {Ishizaki}}\ and\ \bibinfo {author} {\bibfnamefont {H.}~\bibnamefont
  {Nagano}},\ }\href {https://doi.org/10.1016/j.infrared.2019.04.023}
  {\bibfield  {journal} {\bibinfo  {journal} {Infrared Phys. Technol.}\
  }\textbf {\bibinfo {volume} {99}},\ \bibinfo {pages} {248} (\bibinfo {year}
  {2019})}\BibitemShut {NoStop}%
\bibitem [{\citenamefont {Jiang}\ \emph {et~al.}(2018)\citenamefont {Jiang},
  \citenamefont {Qian},\ and\ \citenamefont {Yang}}]{Jiang2018Oct}%
  \BibitemOpen
  \bibfield  {author} {\bibinfo {author} {\bibfnamefont {P.}~\bibnamefont
  {Jiang}}, \bibinfo {author} {\bibfnamefont {X.}~\bibnamefont {Qian}},\ and\
  \bibinfo {author} {\bibfnamefont {R.}~\bibnamefont {Yang}},\ }\bibfield
  {journal} {\bibinfo  {journal} {J. Appl. Phys.}\ }\textbf {\bibinfo {volume}
  {124}},\ \href {https://doi.org/10.1063/1.5046944} {10.1063/1.5046944}
  (\bibinfo {year} {2018})\BibitemShut {NoStop}%
\bibitem [{\citenamefont {Kirsch}\ \emph {et~al.}(2024)\citenamefont {Kirsch},
  \citenamefont {Martin}, \citenamefont {Warzoha}, \citenamefont {McLean},
  \citenamefont {Windover},\ and\ \citenamefont {Takeuchi}}]{Kirsch2024Oct}%
  \BibitemOpen
  \bibfield  {author} {\bibinfo {author} {\bibfnamefont {D.~J.}\ \bibnamefont
  {Kirsch}}, \bibinfo {author} {\bibfnamefont {J.}~\bibnamefont {Martin}},
  \bibinfo {author} {\bibfnamefont {R.}~\bibnamefont {Warzoha}}, \bibinfo
  {author} {\bibfnamefont {M.}~\bibnamefont {McLean}}, \bibinfo {author}
  {\bibfnamefont {D.}~\bibnamefont {Windover}},\ and\ \bibinfo {author}
  {\bibfnamefont {I.}~\bibnamefont {Takeuchi}},\ }\bibfield  {journal}
  {\bibinfo  {journal} {Rev. Sci. Instrum.}\ }\textbf {\bibinfo {volume}
  {95}},\ \href {https://doi.org/10.1063/5.0213738} {10.1063/5.0213738}
  (\bibinfo {year} {2024})\BibitemShut {NoStop}%
\bibitem [{\citenamefont {Cahill}(2004)}]{Cahill2004Dec}%
  \BibitemOpen
  \bibfield  {author} {\bibinfo {author} {\bibfnamefont {D.~G.}\ \bibnamefont
  {Cahill}},\ }\href {https://doi.org/10.1063/1.1819431} {\bibfield  {journal}
  {\bibinfo  {journal} {Rev. Sci. Instrum.}\ }\textbf {\bibinfo {volume}
  {75}},\ \bibinfo {pages} {5119} (\bibinfo {year} {2004})}\BibitemShut
  {NoStop}%
\bibitem [{\citenamefont {Braun}\ and\ \citenamefont
  {Hopkins}(2017)}]{Braun2017May}%
  \BibitemOpen
  \bibfield  {author} {\bibinfo {author} {\bibfnamefont {J.~L.}\ \bibnamefont
  {Braun}}\ and\ \bibinfo {author} {\bibfnamefont {P.~E.}\ \bibnamefont
  {Hopkins}},\ }\bibfield  {journal} {\bibinfo  {journal} {J. Appl. Phys.}\
  }\textbf {\bibinfo {volume} {121}},\ \href
  {https://doi.org/10.1063/1.4982915} {10.1063/1.4982915} (\bibinfo {year}
  {2017})\BibitemShut {NoStop}%
\bibitem [{\citenamefont {Yang}\ \emph
  {et~al.}(2016{\natexlab{a}})\citenamefont {Yang}, \citenamefont {Ziade},\
  and\ \citenamefont {Schmidt}}]{Yang2016Jan}%
  \BibitemOpen
  \bibfield  {author} {\bibinfo {author} {\bibfnamefont {J.}~\bibnamefont
  {Yang}}, \bibinfo {author} {\bibfnamefont {E.}~\bibnamefont {Ziade}},\ and\
  \bibinfo {author} {\bibfnamefont {A.~J.}\ \bibnamefont {Schmidt}},\ }\href
  {https://doi.org/10.1063/1.4939671} {\bibfield  {journal} {\bibinfo
  {journal} {Rev. Sci. Instrum.}\ }\textbf {\bibinfo {volume} {87}},\ \bibinfo
  {pages} {014901} (\bibinfo {year} {2016}{\natexlab{a}})}\BibitemShut
  {NoStop}%
\bibitem [{\citenamefont {Xie}\ \emph {et~al.}(2016)\citenamefont {Xie},
  \citenamefont {Li}, \citenamefont {Tsai}, \citenamefont {Liu}, \citenamefont
  {Braun},\ and\ \citenamefont {Cahill}}]{Xie2016Feb}%
  \BibitemOpen
  \bibfield  {author} {\bibinfo {author} {\bibfnamefont {X.}~\bibnamefont
  {Xie}}, \bibinfo {author} {\bibfnamefont {D.}~\bibnamefont {Li}}, \bibinfo
  {author} {\bibfnamefont {T.-H.}\ \bibnamefont {Tsai}}, \bibinfo {author}
  {\bibfnamefont {J.}~\bibnamefont {Liu}}, \bibinfo {author} {\bibfnamefont
  {P.~V.}\ \bibnamefont {Braun}},\ and\ \bibinfo {author} {\bibfnamefont
  {D.~G.}\ \bibnamefont {Cahill}},\ }\href
  {https://doi.org/10.1021/acs.macromol.5b02477} {\bibfield  {journal}
  {\bibinfo  {journal} {Macromolecules}\ }\textbf {\bibinfo {volume} {49}},\
  \bibinfo {pages} {972} (\bibinfo {year} {2016})}\BibitemShut {NoStop}%
\bibitem [{\citenamefont {Yang}\ \emph
  {et~al.}(2016{\natexlab{b}})\citenamefont {Yang}, \citenamefont {Ziade},\
  and\ \citenamefont {Schmidt}}]{Yang2016Mar}%
  \BibitemOpen
  \bibfield  {author} {\bibinfo {author} {\bibfnamefont {J.}~\bibnamefont
  {Yang}}, \bibinfo {author} {\bibfnamefont {E.}~\bibnamefont {Ziade}},\ and\
  \bibinfo {author} {\bibfnamefont {A.~J.}\ \bibnamefont {Schmidt}},\ }\href
  {https://doi.org/10.1063/1.4943176} {\bibfield  {journal} {\bibinfo
  {journal} {J. Appl. Phys.}\ }\textbf {\bibinfo {volume} {119}},\ \bibinfo
  {pages} {095107} (\bibinfo {year} {2016}{\natexlab{b}})}\BibitemShut
  {NoStop}%
\bibitem [{\citenamefont {Wolf}\ \emph {et~al.}(2004)\citenamefont {Wolf},
  \citenamefont {Pohl},\ and\ \citenamefont {Brendel}}]{Wolf2004Dec}%
  \BibitemOpen
  \bibfield  {author} {\bibinfo {author} {\bibfnamefont {A.}~\bibnamefont
  {Wolf}}, \bibinfo {author} {\bibfnamefont {P.}~\bibnamefont {Pohl}},\ and\
  \bibinfo {author} {\bibfnamefont {R.}~\bibnamefont {Brendel}},\ }\href
  {https://doi.org/10.1063/1.1811390} {\bibfield  {journal} {\bibinfo
  {journal} {J. Appl. Phys.}\ }\textbf {\bibinfo {volume} {96}},\ \bibinfo
  {pages} {6306} (\bibinfo {year} {2004})}\BibitemShut {NoStop}%
\bibitem [{\citenamefont {Scott}\ \emph {et~al.}(2018)\citenamefont {Scott},
  \citenamefont {Smith}, \citenamefont {Henry}, \citenamefont {Rost},
  \citenamefont {Giri}, \citenamefont {Gaskins}, \citenamefont {Fields},
  \citenamefont {Jaszewski}, \citenamefont {Ihlefeld},\ and\ \citenamefont
  {Hopkins}}]{Scott2018Nov}%
  \BibitemOpen
  \bibfield  {author} {\bibinfo {author} {\bibfnamefont {E.~A.}\ \bibnamefont
  {Scott}}, \bibinfo {author} {\bibfnamefont {S.~W.}\ \bibnamefont {Smith}},
  \bibinfo {author} {\bibfnamefont {M.~D.}\ \bibnamefont {Henry}}, \bibinfo
  {author} {\bibfnamefont {C.~M.}\ \bibnamefont {Rost}}, \bibinfo {author}
  {\bibfnamefont {A.}~\bibnamefont {Giri}}, \bibinfo {author} {\bibfnamefont
  {J.~T.}\ \bibnamefont {Gaskins}}, \bibinfo {author} {\bibfnamefont {S.~S.}\
  \bibnamefont {Fields}}, \bibinfo {author} {\bibfnamefont {S.~T.}\
  \bibnamefont {Jaszewski}}, \bibinfo {author} {\bibfnamefont {J.~F.}\
  \bibnamefont {Ihlefeld}},\ and\ \bibinfo {author} {\bibfnamefont {P.~E.}\
  \bibnamefont {Hopkins}},\ }\bibfield  {journal} {\bibinfo  {journal} {Appl.
  Phys. Lett.}\ }\textbf {\bibinfo {volume} {113}},\ \href
  {https://doi.org/10.1063/1.5052244} {10.1063/1.5052244} (\bibinfo {year}
  {2018})\BibitemShut {NoStop}%
\bibitem [{\citenamefont {Ziade}(2020)}]{Ziade2020Dec}%
  \BibitemOpen
  \bibfield  {author} {\bibinfo {author} {\bibfnamefont {E.}~\bibnamefont
  {Ziade}},\ }\bibfield  {journal} {\bibinfo  {journal} {Rev. Sci. Instrum.}\
  }\textbf {\bibinfo {volume} {91}},\ \href {https://doi.org/10.1063/5.0021917}
  {10.1063/5.0021917} (\bibinfo {year} {2020})\BibitemShut {NoStop}%
\bibitem [{\citenamefont {Paterson}\ \emph {et~al.}(2020)\citenamefont
  {Paterson}, \citenamefont {Singhal}, \citenamefont {Tainoff}, \citenamefont
  {Richard},\ and\ \citenamefont {Bourgeois}}]{Paterson2020Jun}%
  \BibitemOpen
  \bibfield  {author} {\bibinfo {author} {\bibfnamefont {J.}~\bibnamefont
  {Paterson}}, \bibinfo {author} {\bibfnamefont {D.}~\bibnamefont {Singhal}},
  \bibinfo {author} {\bibfnamefont {D.}~\bibnamefont {Tainoff}}, \bibinfo
  {author} {\bibfnamefont {J.}~\bibnamefont {Richard}},\ and\ \bibinfo {author}
  {\bibfnamefont {O.}~\bibnamefont {Bourgeois}},\ }\href
  {https://doi.org/10.1063/5.0004576} {\bibfield  {journal} {\bibinfo
  {journal} {J. Appl. Phys.}\ }\textbf {\bibinfo {volume} {127}},\ \bibinfo
  {pages} {245105} (\bibinfo {year} {2020})}\BibitemShut {NoStop}%
\bibitem [{\citenamefont {Liu}\ and\ \citenamefont
  {Asheghi}(2005)}]{Liu2005Jun}%
  \BibitemOpen
  \bibfield  {author} {\bibinfo {author} {\bibfnamefont {W.}~\bibnamefont
  {Liu}}\ and\ \bibinfo {author} {\bibfnamefont {M.}~\bibnamefont {Asheghi}},\
  }\href {https://doi.org/10.1115/1.2130403} {\bibfield  {journal} {\bibinfo
  {journal} {J. Heat Transfer}\ }\textbf {\bibinfo {volume} {128}},\ \bibinfo
  {pages} {75} (\bibinfo {year} {2005})}\BibitemShut {NoStop}%
\bibitem [{\citenamefont {Cancellieri}\ \emph {et~al.}(2020)\citenamefont
  {Cancellieri}, \citenamefont {Scott}, \citenamefont {Braun}, \citenamefont
  {King}, \citenamefont {Oviedo}, \citenamefont {Jezewski}, \citenamefont
  {Richards}, \citenamefont {La~Mattina}, \citenamefont {Jeurgens},\ and\
  \citenamefont {Hopkins}}]{Cancellieri2020Nov}%
  \BibitemOpen
  \bibfield  {author} {\bibinfo {author} {\bibfnamefont {C.}~\bibnamefont
  {Cancellieri}}, \bibinfo {author} {\bibfnamefont {E.~A.}\ \bibnamefont
  {Scott}}, \bibinfo {author} {\bibfnamefont {J.}~\bibnamefont {Braun}},
  \bibinfo {author} {\bibfnamefont {S.~W.}\ \bibnamefont {King}}, \bibinfo
  {author} {\bibfnamefont {R.}~\bibnamefont {Oviedo}}, \bibinfo {author}
  {\bibfnamefont {C.}~\bibnamefont {Jezewski}}, \bibinfo {author}
  {\bibfnamefont {J.}~\bibnamefont {Richards}}, \bibinfo {author}
  {\bibfnamefont {F.}~\bibnamefont {La~Mattina}}, \bibinfo {author}
  {\bibfnamefont {L.~P.~H.}\ \bibnamefont {Jeurgens}},\ and\ \bibinfo {author}
  {\bibfnamefont {P.~E.}\ \bibnamefont {Hopkins}},\ }\bibfield  {journal}
  {\bibinfo  {journal} {J. Appl. Phys.}\ }\textbf {\bibinfo {volume} {128}},\
  \href {https://doi.org/10.1063/5.0019907} {10.1063/5.0019907} (\bibinfo
  {year} {2020})\BibitemShut {NoStop}%
\bibitem [{\citenamefont {Scott}\ \emph {et~al.}(2021)\citenamefont {Scott},
  \citenamefont {Braun}, \citenamefont {Hattar}, \citenamefont {Sugar},
  \citenamefont {Gaskins}, \citenamefont {Goorsky}, \citenamefont {King},\ and\
  \citenamefont {Hopkins}}]{Scott2021Feb}%
  \BibitemOpen
  \bibfield  {author} {\bibinfo {author} {\bibfnamefont {E.~A.}\ \bibnamefont
  {Scott}}, \bibinfo {author} {\bibfnamefont {J.~L.}\ \bibnamefont {Braun}},
  \bibinfo {author} {\bibfnamefont {K.}~\bibnamefont {Hattar}}, \bibinfo
  {author} {\bibfnamefont {J.~D.}\ \bibnamefont {Sugar}}, \bibinfo {author}
  {\bibfnamefont {J.~T.}\ \bibnamefont {Gaskins}}, \bibinfo {author}
  {\bibfnamefont {M.}~\bibnamefont {Goorsky}}, \bibinfo {author} {\bibfnamefont
  {S.~W.}\ \bibnamefont {King}},\ and\ \bibinfo {author} {\bibfnamefont
  {P.~E.}\ \bibnamefont {Hopkins}},\ }\href {https://doi.org/10.1063/5.0038972}
  {\bibfield  {journal} {\bibinfo  {journal} {J. Appl. Phys.}\ }\textbf
  {\bibinfo {volume} {129}},\ \bibinfo {pages} {055307} (\bibinfo {year}
  {2021})}\BibitemShut {NoStop}%
\bibitem [{\citenamefont {Yang}\ \emph {et~al.}(2013)\citenamefont {Yang},
  \citenamefont {Maragliano},\ and\ \citenamefont {Schmidt}}]{Yang2013Oct}%
  \BibitemOpen
  \bibfield  {author} {\bibinfo {author} {\bibfnamefont {J.}~\bibnamefont
  {Yang}}, \bibinfo {author} {\bibfnamefont {C.}~\bibnamefont {Maragliano}},\
  and\ \bibinfo {author} {\bibfnamefont {A.~J.}\ \bibnamefont {Schmidt}},\
  }\href {https://doi.org/10.1063/1.4824143} {\bibfield  {journal} {\bibinfo
  {journal} {Rev. Sci. Instrum.}\ }\textbf {\bibinfo {volume} {84}},\ \bibinfo
  {pages} {104904} (\bibinfo {year} {2013})}\BibitemShut {NoStop}%
\bibitem [{\citenamefont {Mendioroz}\ \emph {et~al.}(2009)\citenamefont
  {Mendioroz}, \citenamefont {Fuente-Dacal}, \citenamefont
  {Api{\ifmmode\tilde{n}\else\~{n}\fi}aniz},\ and\ \citenamefont
  {Salazar}}]{Mendioroz2009Jul}%
  \BibitemOpen
  \bibfield  {author} {\bibinfo {author} {\bibfnamefont {A.}~\bibnamefont
  {Mendioroz}}, \bibinfo {author} {\bibfnamefont {R.}~\bibnamefont
  {Fuente-Dacal}}, \bibinfo {author} {\bibfnamefont {E.}~\bibnamefont
  {Api{\ifmmode\tilde{n}\else\~{n}\fi}aniz}},\ and\ \bibinfo {author}
  {\bibfnamefont {A.}~\bibnamefont {Salazar}},\ }\href
  {https://doi.org/10.1063/1.3176467} {\bibfield  {journal} {\bibinfo
  {journal} {Rev. Sci. Instrum.}\ }\textbf {\bibinfo {volume} {80}},\ \bibinfo
  {pages} {074904} (\bibinfo {year} {2009})}\BibitemShut {NoStop}%
\bibitem [{\citenamefont {Fabbri}\ and\ \citenamefont
  {Fenici}(1995)}]{Fabbri1995Jun}%
  \BibitemOpen
  \bibfield  {author} {\bibinfo {author} {\bibfnamefont {L.}~\bibnamefont
  {Fabbri}}\ and\ \bibinfo {author} {\bibfnamefont {P.}~\bibnamefont
  {Fenici}},\ }\href {https://doi.org/10.1063/1.1146443} {\bibfield  {journal}
  {\bibinfo  {journal} {Rev. Sci. Instrum.}\ }\textbf {\bibinfo {volume}
  {66}},\ \bibinfo {pages} {3593} (\bibinfo {year} {1995})}\BibitemShut
  {NoStop}%
\bibitem [{\citenamefont {Cheaito}\ \emph {et~al.}(2015)\citenamefont
  {Cheaito}, \citenamefont {Gaskins}, \citenamefont {Caplan}, \citenamefont
  {Donovan}, \citenamefont {Foley}, \citenamefont {Giri}, \citenamefont {Duda},
  \citenamefont {Szwejkowski}, \citenamefont {Constantin}, \citenamefont
  {Brown-Shaklee}, \citenamefont {Ihlefeld},\ and\ \citenamefont
  {Hopkins}}]{Cheaito2015Jan}%
  \BibitemOpen
  \bibfield  {author} {\bibinfo {author} {\bibfnamefont {R.}~\bibnamefont
  {Cheaito}}, \bibinfo {author} {\bibfnamefont {J.~T.}\ \bibnamefont
  {Gaskins}}, \bibinfo {author} {\bibfnamefont {M.~E.}\ \bibnamefont {Caplan}},
  \bibinfo {author} {\bibfnamefont {B.~F.}\ \bibnamefont {Donovan}}, \bibinfo
  {author} {\bibfnamefont {B.~M.}\ \bibnamefont {Foley}}, \bibinfo {author}
  {\bibfnamefont {A.}~\bibnamefont {Giri}}, \bibinfo {author} {\bibfnamefont
  {J.~C.}\ \bibnamefont {Duda}}, \bibinfo {author} {\bibfnamefont {C.~J.}\
  \bibnamefont {Szwejkowski}}, \bibinfo {author} {\bibfnamefont
  {C.}~\bibnamefont {Constantin}}, \bibinfo {author} {\bibfnamefont {H.~J.}\
  \bibnamefont {Brown-Shaklee}}, \bibinfo {author} {\bibfnamefont {J.~F.}\
  \bibnamefont {Ihlefeld}},\ and\ \bibinfo {author} {\bibfnamefont {P.~E.}\
  \bibnamefont {Hopkins}},\ }\href {https://doi.org/10.1103/PhysRevB.91.035432}
  {\bibfield  {journal} {\bibinfo  {journal} {Phys. Rev. B}\ }\textbf {\bibinfo
  {volume} {91}},\ \bibinfo {pages} {035432} (\bibinfo {year}
  {2015})}\BibitemShut {NoStop}%
\bibitem [{\citenamefont {Aller}\ \emph {et~al.}(2021)\citenamefont {Aller},
  \citenamefont {Malen},\ and\ \citenamefont {McGaughey}}]{Aller2021Jun}%
  \BibitemOpen
  \bibfield  {author} {\bibinfo {author} {\bibfnamefont {H.~T.}\ \bibnamefont
  {Aller}}, \bibinfo {author} {\bibfnamefont {J.~A.}\ \bibnamefont {Malen}},\
  and\ \bibinfo {author} {\bibfnamefont {A.~J.~H.}\ \bibnamefont {McGaughey}},\
  }\href {https://doi.org/10.1103/PhysRevApplied.15.064043} {\bibfield
  {journal} {\bibinfo  {journal} {Phys. Rev. Appl.}\ }\textbf {\bibinfo
  {volume} {15}},\ \bibinfo {pages} {064043} (\bibinfo {year}
  {2021})}\BibitemShut {NoStop}%
\bibitem [{\citenamefont {Monachon}\ \emph {et~al.}(2016)\citenamefont
  {Monachon}, \citenamefont {Weber},\ and\ \citenamefont
  {Dames}}]{Monachon2016Jul}%
  \BibitemOpen
  \bibfield  {author} {\bibinfo {author} {\bibfnamefont {C.}~\bibnamefont
  {Monachon}}, \bibinfo {author} {\bibfnamefont {L.}~\bibnamefont {Weber}},\
  and\ \bibinfo {author} {\bibfnamefont {C.}~\bibnamefont {Dames}},\ }\href
  {https://doi.org/10.1146/annurev-matsci-070115-031719} {\bibfield  {journal}
  {\bibinfo  {journal} {Annu. Rev. Mater. Res.}\ ,\ \bibinfo {pages} {433}}
  (\bibinfo {year} {2016})}\BibitemShut {NoStop}%
\bibitem [{\citenamefont {Jiang}\ \emph {et~al.}(2023)\citenamefont {Jiang},
  \citenamefont {Ryu}, \citenamefont {Pachauri}, \citenamefont {Ingebrandt},
  \citenamefont {Vu},\ and\ \citenamefont {Morikawa}}]{Jiang2023Sep}%
  \BibitemOpen
  \bibfield  {author} {\bibinfo {author} {\bibfnamefont {F.}~\bibnamefont
  {Jiang}}, \bibinfo {author} {\bibfnamefont {M.}~\bibnamefont {Ryu}}, \bibinfo
  {author} {\bibfnamefont {V.}~\bibnamefont {Pachauri}}, \bibinfo {author}
  {\bibfnamefont {S.}~\bibnamefont {Ingebrandt}}, \bibinfo {author}
  {\bibfnamefont {X.~T.}\ \bibnamefont {Vu}},\ and\ \bibinfo {author}
  {\bibfnamefont {J.}~\bibnamefont {Morikawa}},\ }\href
  {https://doi.org/10.1063/5.0160602} {\bibfield  {journal} {\bibinfo
  {journal} {Rev. Sci. Instrum.}\ }\textbf {\bibinfo {volume} {94}},\ \bibinfo
  {pages} {094903} (\bibinfo {year} {2023})}\BibitemShut {NoStop}%
\bibitem [{\citenamefont {Alasli}\ \emph {et~al.}(2025)\citenamefont {Alasli},
  \citenamefont {Iguchi}, \citenamefont {Uchida},\ and\ \citenamefont
  {Nagano}}]{Alasli2025Jan}%
  \BibitemOpen
  \bibfield  {author} {\bibinfo {author} {\bibfnamefont {A.}~\bibnamefont
  {Alasli}}, \bibinfo {author} {\bibfnamefont {R.}~\bibnamefont {Iguchi}},
  \bibinfo {author} {\bibfnamefont {K.-i.}\ \bibnamefont {Uchida}},\ and\
  \bibinfo {author} {\bibfnamefont {H.}~\bibnamefont {Nagano}},\ }\href
  {https://doi.org/10.1063/5.0245566} {\bibfield  {journal} {\bibinfo
  {journal} {Appl. Phys. Lett.}\ }\textbf {\bibinfo {volume} {126}},\ \bibinfo
  {pages} {044102} (\bibinfo {year} {2025})}\BibitemShut {NoStop}%
\end{thebibliography}%

\end{document}